\documentclass{ws-ijmpa}
\usepackage[super,compress]{cite}
\usepackage{graphicx}
\usepackage{color}
\begin{document}
\markboth{H. Suganuma, H. Ohata, M. Kitazawa}{Thermodynamic Potential of the Polyakov Loop in SU(3) Quenched Lattice QCD}

%
\catchline{}{}{}{}{}
%

\title{Thermodynamic Potential of the Polyakov Loop \\ 
in SU(3) Quenched Lattice QCD}

\author{Hideo Suganuma}
\address{Department of Physics, Kyoto University\\
Kitashirakawaoiwake, Sakyo, Kyoto 606-8502, Japan\\
suganuma@scphys.kyoto-u.ac.jp}

\author{Hiroki Ohata}
\address{Yukawa Institute for Theoretical Physics (YITP), Kyoto University\\
Kitashirakawaoiwake, Sakyo, Kyoto 606-8502, Japan\\
hiroki.ohata@yukawa.kyoto-u.ac.jp}

\author{Masakiyo Kitazawa}
\address{Yukawa Institute for Theoretical Physics (YITP), Kyoto University\\
Kitashirakawaoiwake, Sakyo, Kyoto 606-8502, Japan\\
masakiyo.kitazawa@yukawa.kyoto-u.ac.jp}

\maketitle

\begin{history}
\end{history}

\begin{abstract}
Using SU(3) lattice QCD, we study for the first time the effective potential of the Polyakov loop $\langle P \rangle$ at finite temperature, i.e., the thermodynamic potential, in the field-theoretical way. 
In the framework of the reweighting method in lattice QCD, 
we express the effective potential 
$V_{\rm eff}(\langle P \rangle)$
using the expectation value without a source term.  
In particular, we consider the most difficult and interesting case 
of vacuum coexistence at the critical temperature $T_c$. 
We adopt SU(3) quenched lattice QCD on 
$48^3 \times 6$ at $\beta$= 5.89379, which corresponds exactly  
to the critical temperature $T_c$ of the deconfinement phase transition, 
and use 200,000 Monte Carlo configurations. 
After categorizing the gauge configurations into 
one $Z_3$-symmetric and three $Z_3$-broken vacua each, 
we perform a vacuum-associated reweighting method, 
using the gauge configurations around each vacuum separately. 
Finally, we obtain the Polyakov-loop effective potential, which is well depicted around the $Z_3$-symmetric and $Z_3$-broken vacua.

\end{abstract}

\ccode{PACS numbers:12.38.Gc, 12.38.Mh, 12.38.Aw}


\section{Introduction}

Quantum chromodynamics (QCD) is 
the fundamental theory of the strong interaction and a
major component of the Standard Model 
in elementary particle physics \cite{GWP73}.
While high-energy hadron reactions can be treated as perturbative QCD, 
low-energy QCD is strongly coupled and has a complicated vacuum, 
which exhibits color confinement 
and spontaneous chiral symmetry breaking 
as the most striking phenomena of 
nonperturbative QCD~\cite{Polyakov78,Creutz80,Rothe2012}. 
In particular, color confinement is a very curious phenomenon 
peculiar to QCD~\cite{Handbook2023}, and its physical understanding is still 
an extremely difficult problem. 

Finite-temperature QCD is also an interesting topic in particle physics. 
In fact, quark-gluon plasma (QGP) 
has been well studied 
both theoretically and experimentally, 
and the thermal QCD transition is 
also important to obtain the physical description of the early universe.\cite{YHM2005} 

In general, to describe the thermal system and 
the phase transition, it is useful to study the thermodynamic potential, which in field theory is formulated as the effective potential at the finite temperature. 
For example, the effective potential indicates the order of the phase transition, and, in the case of the first-order transition, 
the vacuum-to-vacuum transition rate at the critical temperature.  
From the effective potential at the critical temperature 
$T_c$, the thermal transition probability between the confined and deconfined vacua can be estimated quantitatively, and so on. 
However, the Polyakov-loop effective potential has not yet been calculated directly 
in lattice QCD within the quantum field theory, although some attempts have been 
made using the strong-coupling expansion of QCD\cite{FLLP2011,FS2017}.


In this paper, we study for the first time 
the thermodynamic potential of a confinement order parameter, 
i.e., the effective potential of the Polyakov loop 
at finite temperature, in SU(3) lattice QCD 
in a field-theoretical way.
To focus on quark confinement, we demonstrate it in quenched QCD 
without dynamical quarks, although the present formalism 
also applies to full QCD. 

\section{Quark confinement, Polyakov loop and $Z_{N_c}$ center symmetry}

In thermal QCD, 
a standard order parameter of 
the quark confinement is 
the Polyakov loop \cite{Polyakov78}, which is 
defined as a path-ordered product 
along the imaginary time,
\begin{eqnarray}
P({\bf x}) \equiv \frac{1}{N_c}{\rm Tr}~{\rm P} \exp\{i g\int_0^{1/T}dt A_t({\bf x}, t)\},
\end{eqnarray}
at finite temperature $T$. 
Here, $g$ is the QCD gauge coupling and P denotes the path-ordered product.
The thermal expectation 
value of the Polyakov loop 
$\langle P \rangle$ 
is related to the single-quark free energy $E_q$ as
$\langle P \rangle \propto e^{-E_q/T}$.
In the confinement phase, 
the Polyakov loop $\langle P \rangle$ is (almost) zero, 
because the single-quark energy is infinite $E_q=\infty$ (or large), 
reflecting the quark confinement.
In the deconfinement phase, the Polyakov loop $\langle P \rangle$ takes a 
finite value, 
and the single-quark free energy $E_q$ becomes finite.
The Polyakov loop $\langle P \rangle$ 
thus behaves as an 
order parameter of the quark confinement \cite{Polyakov78,Rothe2012}.


The Polyakov loop is also an order parameter of 
the $Z_{N_c}$ center symmetry \cite{Polyakov78,Rothe2012}.
Considering lattice QCD defined on a periodic 
$L_s^3 \times L_t$ lattice with spacing $a$, 
the gauge variables are described by link-variables 
$U_\mu(x)=e^{iagA_\mu(x)}\in {\rm SU}(N_c)$,
and the Polyakov loop $P$ is expressed as a product of 
the temporal link-variables $U_4({\bf x},t)$,
\begin{eqnarray}
P({\bf x}) \equiv \frac{1}{N_c}{\rm Tr} ~ [U_4({\bf x},a) U_4({\bf x},2a)
\cdots U_4({\bf x},L_ta)]
= P_{\rm Re}({\bf x})+iP_{\rm Im}({\bf x}) \in {\bf C}.
\end{eqnarray}
The $Z_{N_c}$ center symmetry relates to 
the spatially global transformation 
of the temporal link-variables at a fixed time $t$, 
$U_4({\bf x},t) \rightarrow z U_4({\bf x},t)$, 
with a center element $z \in Z_{N_c}$ of SU($N_c$).
Under the $Z_{N_c}$ transformation, 
while the lattice gauge action is invariant, 
the Polyakov loop $P$ is variant 
and $\langle P \rangle$
behaves as an order parameter of the $Z_{N_c}$ center symmetry. 
Indeed, the confinement phase with $\langle P \rangle = 0$ 
is $Z_{N_c}$-symmetric, and the $Z_{N_c}$ center symmetry is 
spontaneously broken in the deconfinement phase 
with $\langle P \rangle \ne 0$. 


Despite the importance of the Polyakov loop $\langle P \rangle$ for confinement properties in QCD, its thermodynamic potential has not been directly calculated in lattice QCD.

\section{Derivation of the effective potential of the Polyakov loop in lattice QCD}

We derive the effective potential (thermodynamic potential) 
$V_{\rm eff}(\langle P \rangle)$ of the Polyakov loop $\langle P \rangle$ using the standard field theory method based on lattice formalism.

\subsection{Field-theoretical derivation 
of Polyakov-loop effective potential}

The effective potential $V_{\rm eff}(\langle P \rangle)$ physically means 
the (free) energy of the vacuum specified with an arbitrary value 
of $\langle P \rangle \in {\bf C}$, 
which is not restricted to the ground-state value. 
In general, for the treatment of various  systems beyond 
the ground-state vacuum, 
it is necessary to introduce the source term  
in order to modify the stable point 
in a field-theoretical way~\cite{H92,R96}.

In this case, we introduce the source $J$ of the Polyakov loop, 
\begin{eqnarray}
J({\bf x})= J_{\rm Re}({\bf x})+iJ_{\rm Im}({\bf x}) \in {\bf C}, 
\end{eqnarray}
and define the QCD generating functional including 
the Polyakov-loop source term, 
\begin{eqnarray}
Z[J]&=&\int DU \exp[-\{S[U] + {\rm Re}(J \cdot P[U]^*)\}] \cr 
&=&\int DU \exp[-\{S[U] +(J_{\rm Re} \cdot P_{\rm Re}[U]+J_{\rm Im} \cdot P_{\rm Im}[U])\}]. 
\end{eqnarray}
Here, the inner product denotes 
the spatial summation 
on the lattice such as  
\begin{eqnarray}
J \cdot P^* \equiv 
\sum_{\bf x}J({\bf x}) \cdot P({\bf x})^*.
\end{eqnarray}
In this paper, we are mainly concerned with the constant source of 
$J({\bf x})=J$, which  
preserves the translational symmetry of the system. 

The generating functional $W[J]$ for connected diagrams 
is defined as usual~\cite{H92,R96},
\begin{eqnarray}
W[J] \equiv -\ln Z[J].
\end{eqnarray}
The derivative of 
$W[J]$ with respect to the source $J$ 
gives the expectation value of the Polyakov loop $\langle P \rangle_J$ in the presence of the source $J$:
\begin{eqnarray}
\frac{\partial W[J_{\rm Re}, J_{\rm Im}]}{\partial J_{\rm Re}}&=&
\frac{1}{Z[J]}\int DU e^{-(S+J_{\rm Re}\cdot P_{\rm Re}+J_{\rm Im}\cdot P_{\rm Im})} P_{\rm Re}=\langle P_{\rm Re} \rangle_J,
\cr
\cr
\frac{\partial W[J_{\rm Re}, J_{\rm Im}]}{\partial J_{\rm Im}}&=&
\frac{1}{Z[J]}\int DU e^{-(S+J_{\rm Re}\cdot P_{\rm Re}+J_{\rm Im}\cdot P_{\rm Im})} P_{\rm Im}=\langle P_{\rm Im} \rangle_J.
\end{eqnarray}
Thus, by 
introducing the source $J$, 
one can treat various systems with 
different values of $\langle P \rangle_J$. 
Note that the Polyakov loop $\langle P \rangle_J$ 
is expressed with the source $J$, 
and then $J$ is also described with the Polyakov loop $\langle P \rangle$ as 
$J[\langle P \rangle]$ in principle.

The effective action 
$\Gamma [\langle P \rangle]$ of 
the Polyakov loop $\langle P \rangle$ is 
defined as its Legendre transformation 
in a standard field-theoretical way\cite{H92,R96},
\begin{eqnarray}
\Gamma[\langle P \rangle]
\equiv W[J]- {\rm Re}(J \cdot \langle P \rangle^*)
=W[J]-(J_{\rm Re} \cdot \langle P_{\rm Re} \rangle
+J_{\rm Im} \cdot \langle P_{\rm Im} \rangle).
\end{eqnarray}
The effective action  
$\Gamma[\langle P \rangle]$ 
can be expressed as a function of various $\langle P \rangle$ using 
the relation of $J[\langle P \rangle]$. 
The effective action $\Gamma[\langle P \rangle]$ 
satisfies the extreme conditions, 
\begin{eqnarray}
\frac{\partial \Gamma [\langle P_{\rm Re} \rangle, \langle P_{\rm Im} \rangle]}{\partial \langle P_{\rm Re} \rangle}=-J_{\rm Re},
\quad 
\frac{\partial \Gamma[\langle P_{\rm Re} \rangle, \langle P_{\rm Im} \rangle]}{\partial \langle P_{\rm Im} \rangle}=-J_{\rm Im}, 
\end{eqnarray}
which lead to the relation $J[\langle P\rangle]$ again.

For the constant source of $J({\bf x})=J$, which preserves translational invariance, 
the effective potential is introduced 
in the Euclidean metric as
\begin{eqnarray}
V_{\rm eff}(\langle P \rangle) \equiv \frac{\Gamma(\langle P \rangle)}{\int _0^{1/T}dt \int d^3x}
=\frac{T}{V}~\Gamma(\langle P \rangle)
=\frac{1}{L_t L_s^3 a^4}\Gamma(\langle P \rangle)
\end{eqnarray}
with the spatial volume $V$ at the temperature $T=1/(L_t a)$, 
which appears as the inverse of the 
imaginary-time period in lattice QCD.
At finite temperatures, the effective potential $V_{\rm eff}[\langle P \rangle]$ physically means 
the thermodynamic potential or the free energy as a function of the Polyakov loop $\langle P \rangle$.
Similar to the effective action, 
the effective potential $V_{\rm eff}[\langle P \rangle]$ 
satisfies the extreme conditions, 
\begin{eqnarray}
\frac{\partial V_{\rm eff}[\langle P_{\rm Re} \rangle, \langle P_{\rm Im} \rangle]}{\partial \langle P_{\rm Re} \rangle}
=-\frac{T}{V}J_{\rm Re},
\quad 
\frac{\partial V_{\rm eff}[\langle P_{\rm Re} \rangle, \langle P_{\rm Im} \rangle]}{\partial \langle P_{\rm Im} \rangle}
=-\frac{T}{V}J_{\rm Im},
\end{eqnarray}
and the stable thermal system is realized at  
the minimum (or extremal) point of the effective potential in the absence of the source, $J=0$.


\subsection{Reweighting method 
to describe with quantities at $J=0$}

In the lattice QCD Monte Carlo method,  
gauge configurations $\{U_n(s)\}_{n=1, 2, \cdots, N_{\rm conf}}$ 
are generated numerically with the most importance sampling,\cite{Rothe2012,Creutz80} 
and the expectation value of any physical quantity is calculated 
as the ensemble average over the $N_{\rm conf}$ lattice gauge configurations,
\begin{eqnarray}
\langle O[U]\rangle=\frac{1}{N_{\rm conf}}\sum_{n=1}^{N_{\rm conf}} O[U_n] +O(N_{\rm conf}^{-1/2}).
\end{eqnarray}
In ordinary lattice QCD, the source term is not introduced 
in the Monte Carlo calculations, 
and the gauge configurations are generated at $J=0$. 


In order to obtain the effective potential $V_{\rm eff}[\langle P\rangle]$, we have to deal with various systems specified by $\langle P \rangle_J$, 
corresponding to the ground state in the presence of the source $J$.
One straightforward way to introduce the source 
is to perform the Monte Carlo calculation  
with the action factor $e^{-[S+{\rm Re}(J\cdot P^*)]}$
for each value of $J$, but this is a rather tedious task.

A more sophisticated way is to use 
the reweighting method~\cite{FS8889, WHOT2016, WHOT2021}, 
where the source factor $e^{-{\rm Re}(J\cdot P^*)}$ 
is treated as an additional operator. 
Using this method, 
the expectation value of operators with $J\ne 0$ 
can be expressed in terms of quantities at $J=0$. 
The expectation values at $J=0$ can be calculated with the gauge configurations 
generated in ordinary lattice QCD.



For example, $W[J]$ can be expressed 
by an expectation value at $J$ = 0, 
\begin{eqnarray}
W[J]-W[J=0]&=&-\ln \frac{Z[J]}{Z[J=0]}
=-\ln \frac{\int DU e^{-S[U]-(J_{\rm Re} \cdot P_{\rm Re}[U]+J_{\rm Im}\cdot P_{\rm Im}[U])}}{\int DU e^{-S[U]}} \cr
\cr
&=&-\ln \langle e^{-(J_{\rm Re} \cdot P_{\rm Im}[U]+J_{\rm Im}\cdot P_{\rm Im}[U])} \rangle_{J=0}, 
\label{eq:W}
\end{eqnarray}
because of 
\begin{eqnarray}
\langle O[U] \rangle_{J=0} =
\frac{\int DU e^{-S[U]}O[U]}{\int DU e^{-S[U]}}. 
\end{eqnarray}
Here, we have added 
an irrelevant $J$-independent constant, $W[J=0]$.

The Polyakov loop $\langle P \rangle_J$ in the presence of the source $J$ 
can also be expressed with expectation values at $J$ = 0:
\begin{eqnarray}
\langle P_{\rm Re/Im}[U] \rangle_J
&=&\frac{\int DU e^{-S[U]-(J_{\rm Re} \cdot P_{\rm Re}[U]+J_{\rm Im}\cdot P_{\rm Im}[U])}
P_{\rm Re/Im}[U]}
{\int DU e^{-S[U]-(J_{\rm Re} \cdot P_{\rm Re}[U]+J_{\rm Im}\cdot P_{\rm Im}[U])}}
\cr
\cr
&=&\frac{\langle e^{-(J_{\rm Re} \cdot P_{\rm Re}[U]+J_{\rm Im}\cdot P_{\rm Im}[U])}P_{\rm Re/Im}[U]\rangle_{J=0}}{\langle e^{-(J_{\rm Re} \cdot P_{\rm Re}[U]+J_{\rm Im}\cdot P_{\rm Im}[U])}\rangle_{J=0}}.
\label{eq:P}
\end{eqnarray}

Thus, physical quantities in the presence of source $J$ can be calculated using expectation values at $J$ = 0, by including the source factor $\exp\{-{\rm Re}~(J \cdot P^*)\}$ 
as an additional operator.

\subsection{Lattice QCD derivation of the effective potential} 

Now, we derive the effective potential 
$V_{\rm eff}[\langle P\rangle]$ 
in lattice QCD 
using the reweighting method.
As mentioned above, $W[J]$ and the Polyakov loop 
$\langle P \rangle_J$ in the presence of the source $J$ 
can be calculated with expectation values at $J$ = 0, 
i.e., Eqs.~(\ref{eq:W}) and (\ref{eq:P}).

Therefore, for the constant source 
$J=J_{\rm Re}+iJ_{\rm Im} \in {\bf C}$, 
we obtain the effective potential 
$V_{\rm eff}[\langle P\rangle]= \frac{T}{V} \Gamma[\langle P \rangle] $
of the Polyakov loop $\langle P \rangle$ 
in the Euclidean metric, 
except for an irrelevant constant:
\begin{eqnarray}
&& \frac{V}{T} V_{\rm eff}[\langle P \rangle]
=(W[J]-W[0]) 
-(J_{\rm Re} \langle P_{\rm Re}[U] \rangle_J
+J_{\rm Im} \langle P_{\rm Im}[U] \rangle_J)
\cr
\cr
&&=-\ln \langle e^{-(J_{\rm Re} \cdot P_{\rm Im}[U]+J_{\rm Im}\cdot P_{\rm Im}[U])} \rangle_{J=0}
\\
&&-J_{\rm Re} 
\frac{\langle e^{-(J_{\rm Re} \cdot P_{\rm Re}[U]+J_{\rm Im}\cdot P_{\rm Im}[U])}P_{\rm Re}\rangle_{J=0}}{\langle e^{-(J_{\rm Re} \cdot P_{\rm Re}[U]+J_{\rm Im}\cdot P_{\rm Im}[U])}\rangle_{J=0}}
-J_{\rm Im} \frac{\langle e^{-(J_{\rm Re} \cdot P_{\rm Re}[U]+J_{\rm Im}\cdot P_{\rm Im}[U])}P_{\rm Im}[U]\rangle_{J=0}}{\langle e^{-(J_{\rm Re} \cdot P_{\rm Re}[U]+J_{\rm Im}\cdot P_{\rm Im}[U])}\rangle_{J=0}}.
\nonumber
\label{eq:ep}
\end{eqnarray}
The expectation values without source ($J$ = 0)
are calculated directly as the ensemble average 
over $N_{\rm conf}$ gauge configurations generated in lattice QCD,
\begin{eqnarray}
\langle e^{-(J_{\rm Re} \cdot P_{\rm Re}[U]+J_{\rm Im}\cdot P_{\rm Im}[U])}\rangle_{J=0}
&=&\frac{1}{N_{\rm conf}}\sum_{n=1}^{N_{\rm conf}}
e^{-(J_{\rm Re} \cdot P_{\rm Re}[U_n]+J_{\rm Im}\cdot P_{\rm Im}[U_n])},
\\ \nonumber
\langle e^{-(J_{\rm Re} \cdot P_{\rm Re}+J_{\rm Im}\cdot P_{\rm Im})} P_{\rm Re/Im}\rangle_{J=0} 
&=&\frac{1}{N_{\rm conf}}\sum_{n=1}^{N_{\rm conf}}
e^{-(J_{\rm Re} \cdot P_{\rm Re}[U_n]+J_{\rm Im}\cdot P_{\rm Im}[U_n])} P_{\rm Re/Im}[U_n].
\end{eqnarray}


Thus, from the gauge configurations generated in lattice QCD at $J=0$, 
for arbitrary $J=J_{\rm Re}+iJ_{\rm Im} \in {\bf C}$, we can calculate 
$\langle P \rangle_J$ and $V_{\rm eff}[\langle P \rangle_J]$
by rewriting them with some expectation values at $J=0$. 

\subsection{Lattice QCD setup and 
caution in describing coexisting vacua}

In this paper, we deal with the most difficult and interesting case 
of vacuum coexistence at the critical temperature $T_c$, 
using SU(3) quenched lattice QCD with the standard plaquette action 
on $48^3 \times 6$ at $\beta$ = 5.89379,~\cite{WHOT2016,WHOT2021}
which corresponds to the critical temperature $T_c \simeq$ 280~MeV 
of the deconfinement phase transition in quenched QCD.
The lattice spacing is $a \simeq$ 0.12~fm, 
and the lattice unit of $a=1$ is used.

We use a large number of 200,000 gauge configurations, which were 
generated in Ref.\cite{WHOT2016} with the pseudo-heat-bath algorithm 
in the ordinary Monte Carlo method at $J=0$.
%
%
As for the Polyakov-loop source 
$J=J_{\rm Re}+i J_{\rm Im} \in {\bf C}$, 
we use 901 different points.  

\begin{figure}[htbp]
    \centering
    \includegraphics[width=8.5cm]{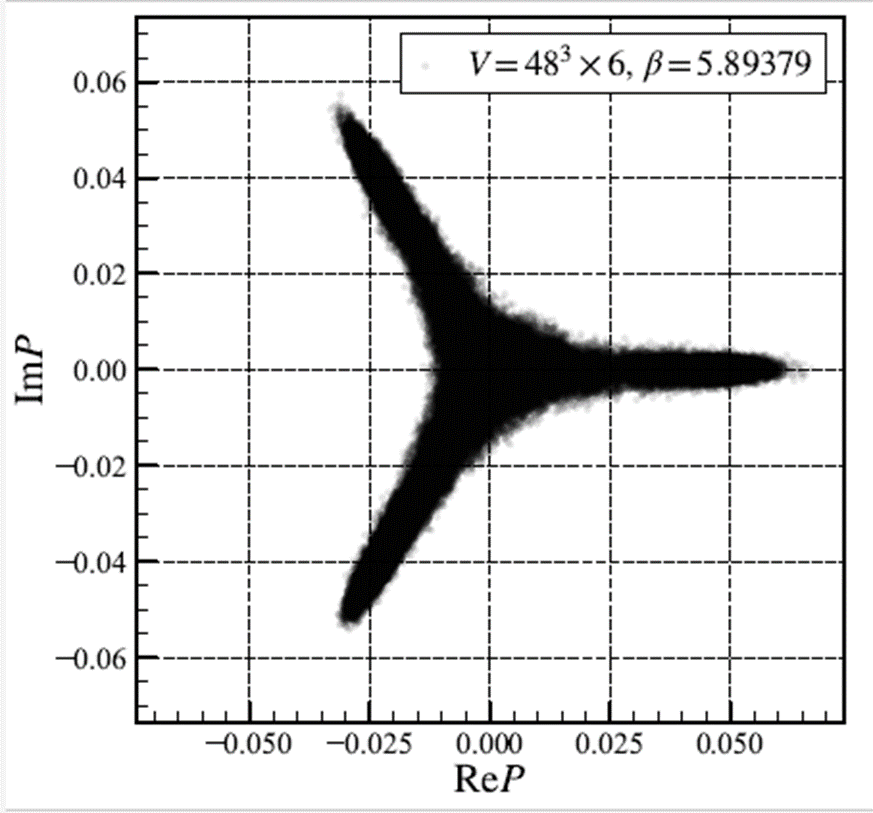}
    \caption{Scatter plot of the Polyakov loop $\langle P \rangle 
    ={\rm Re} \langle P \rangle+i~{\rm Im} \langle P \rangle \in {\bf C}$ 
    at the critical temperature $T_c$.
    The dots correspond to 
    200,000 gauge configurations
    generated in SU(3) quenched lattice QCD 
    on $48^3 \times 6$ at $\beta$ = 5.89379. 
    }
    \label{fig:critical}
\end{figure}

Figure~\ref{fig:critical} shows the scatter plot of the Polyakov loop $\langle P \rangle_{J=0} \in {\bf C}$ 
at the critical temperature $T_c$, where confined and deconfined vacua coexist.
This scatter plot clearly shows the $Z_3$ center symmetry. 
As a note of caution, naive averaging over all the lattice data leads to 
a nonsensical zero expectation value of the Polyakov loop.

\begin{figure}
    \centering
    \includegraphics[width=6.9cm]{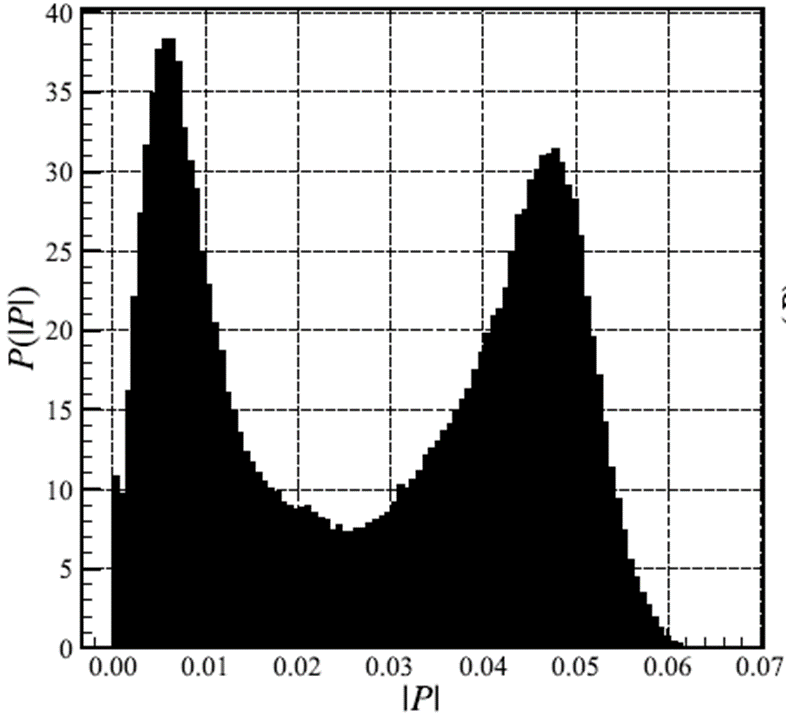}
    \caption{Histogram of the spatially-averaged 
Polyakov loop $|\langle P \rangle|$ of 
each gauge configuration 
at the critical temperature $T_c$. 
The two peaks indicate the coexistence of confined and deconfined vacua at the critical temperature $T_c$.}
    \label{fig:histgram}
\end{figure}

Figure~\ref{fig:histgram} shows the histogram of 
the spatially-averaged 
Polyakov loop $|\langle P \rangle|$ of 
each gauge configuration 
at the critical temperature $T_c$. 
The two peaks indicate the coexistence of confined and deconfined vacua at the critical temperature $T_c$.

Such vacuum coexistence sometimes occurs in thermal QCD. 
In fact, $Z_3$-symmetric and $Z_3$-broken vacua coexist at $T_c$, 
and three $Z_3$-broken vacua coexist above $T_c$.

In general, when vacua coexist, 
there is some caution in describing 
the effective potential or 
the thermodynamic potential. 
For example, even if spontaneous symmetry breaking occurs in the thermodynamic limit,  
naive averaging over all configurations 
in a finite volume 
leads to a zero expectation value of 
an order parameter, which is only an artifact. 
Therefore, naive averaging over 
all the vacua must be avoided 
in the case of coexistence of vacua.

\section{Vacuum-associated reweighting method and 
lattice QCD results} 

As a technical improvement, 
we perform the vacuum-associated reweighting method 
where the gauge configurations around each vacuum 
are used separately, 
after categorizing the gauge configurations into 
each of the $Z_3$-symmetric and three $Z_3$-broken vacua.

\begin{figure}[htbp]
    \centering
    \includegraphics[width=4.15cm]{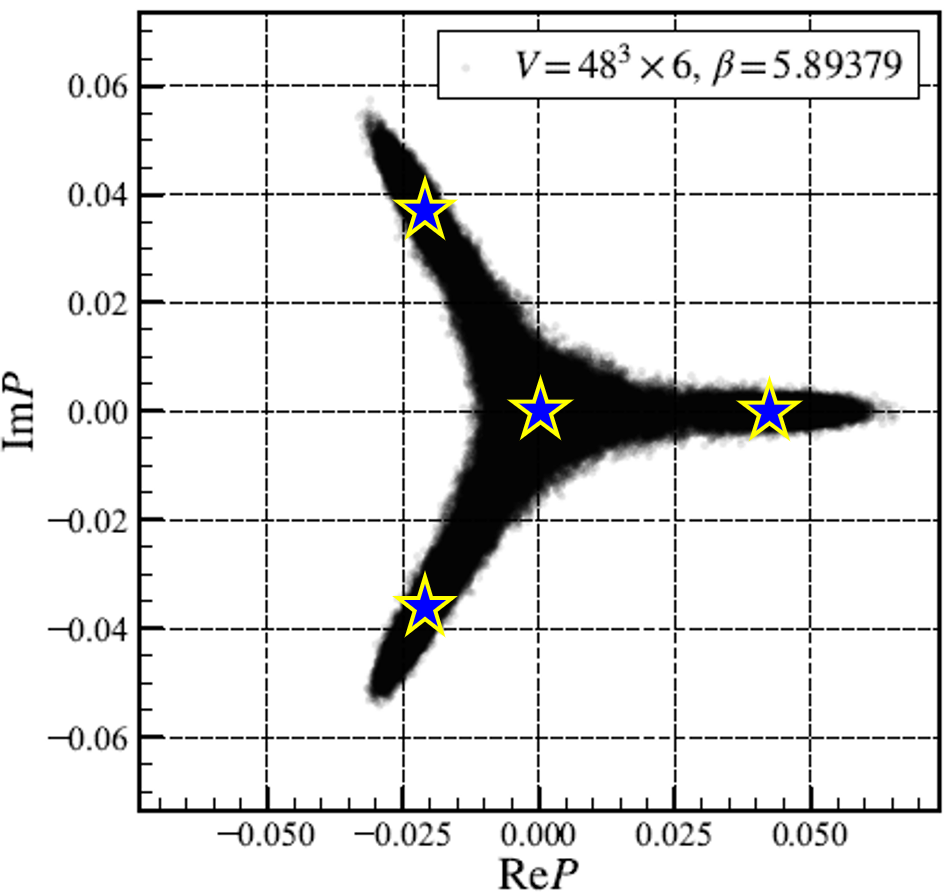}
    \includegraphics[width=4.15cm]{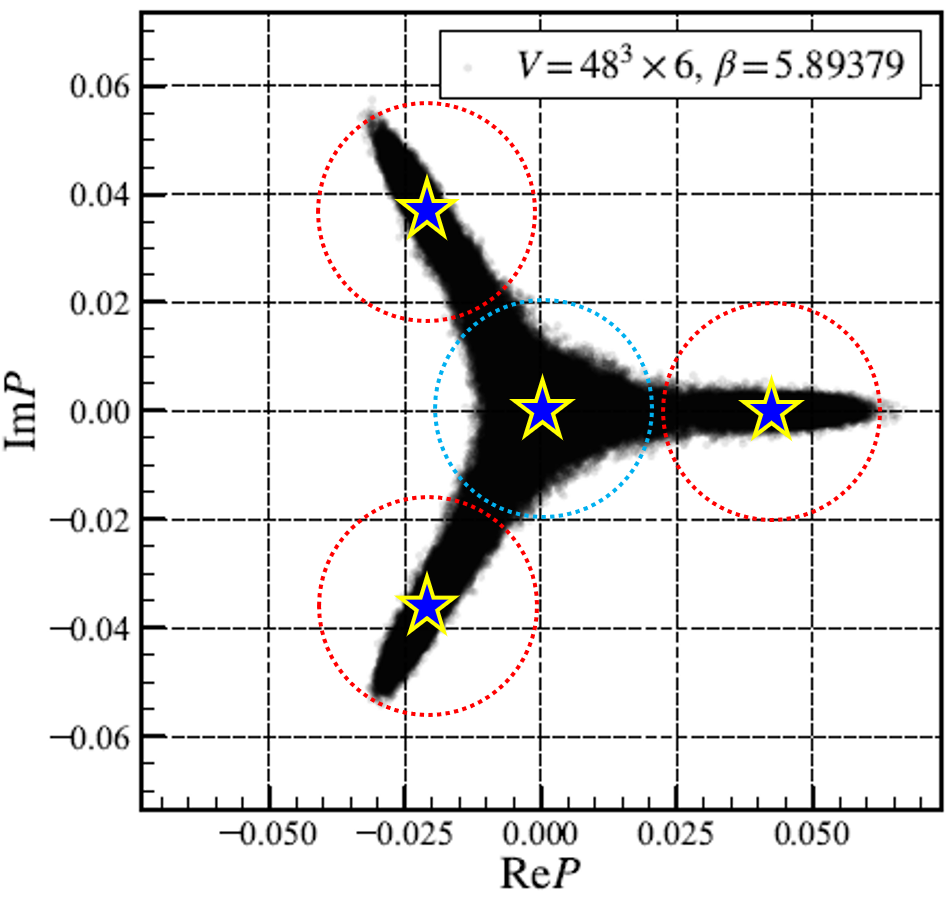}
    \includegraphics[width=4.15cm]{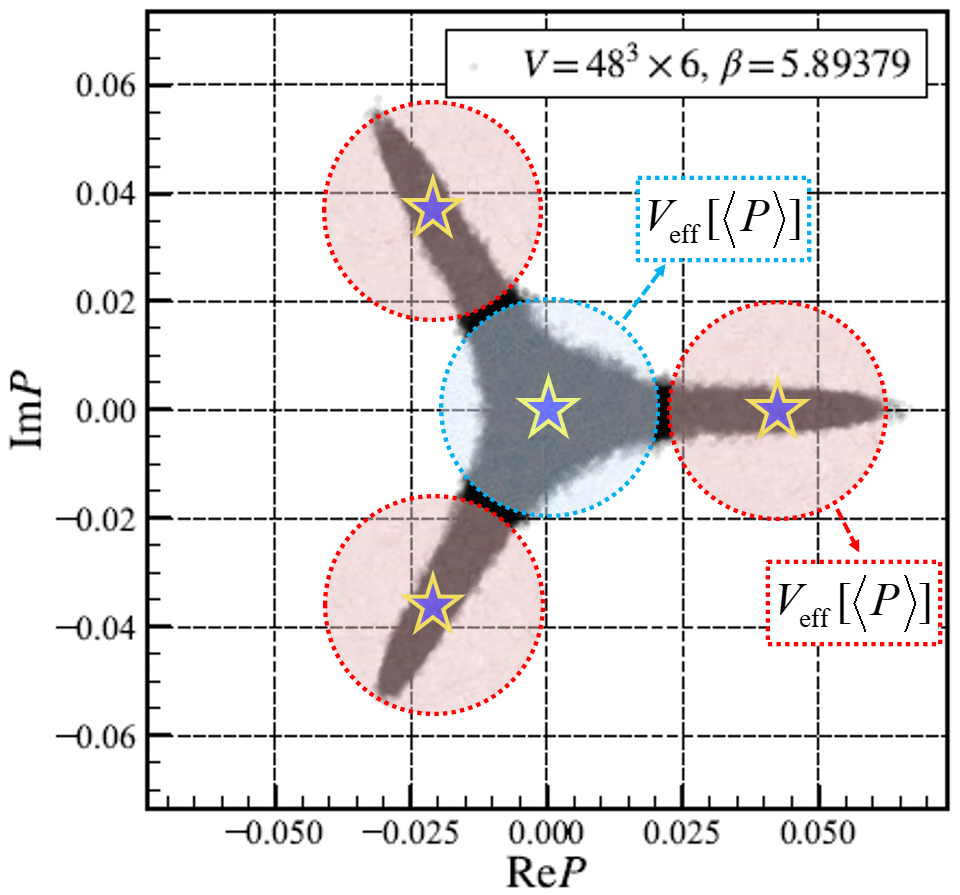}
    \caption{
Procedure of the vacuum-associated reweighting method: 
(Left) For the scatter plot of the Polyakov loop 
$\langle P \rangle$, the $Z_3$-symmetric 
and three $Z_3$-broken vacua are marked with stars.
(Middle) Each lattice gauge configuration is 
associated with the closest vacuum 
among the $Z_3$-symmetric and three $Z_3$-broken vacua.
The area of each vacuum is indicated by the circle.
(Right) Around each vacuum, 
the effective potential $V_{\rm eff}[\langle P \rangle]$ of the Polyakov loop $\langle P \rangle$ is calculated numerically with 
the reweighting method using the lattice configurations 
inside the circle.
}
    \label{fig:procedure}
\end{figure}

Using the vacuum-associated reweighting method, 
the Polyakov-loop effective potential $V_{\rm eff}[\langle P \rangle]$ can be calculated. 
%
%
The specific procedure is as follows:
\begin{enumerate}
\item 
For the scatter plot of the Polyakov loop $\langle P \rangle$ for lattice gauge configurations 
in Fig.~\ref{fig:critical}, 
the $Z_3$-symmetric and three $Z_3$-broken vacua are marked with stars, as shown in Fig.~\ref{fig:procedure}~(left).
\item 
Each lattice gauge configuration 
is associated with the closest vacuum among 
the $Z_3$-symmetric 
and three $Z_3$-broken vacua, as shown in Fig.~\ref{fig:procedure}~(middle). 
\item 
Around each vacuum, we calculate 
the effective potential $V_{\rm eff}[\langle P \rangle]$ 
using the reweighting method with 
the lattice gauge configurations associated with the vacuum, 
as illustrated in Fig.~\ref{fig:procedure}~(right). 
\end{enumerate}
%
In this way, we numerically obtain 
the effective potential (thermodynamic potential, free energy)
$V_{\rm eff}(\langle P \rangle)$ of the Polyakov loop $\langle P \rangle$
in SU(3) lattice QCD. 

\begin{figure}[htbp]
    \includegraphics[width=12cm]{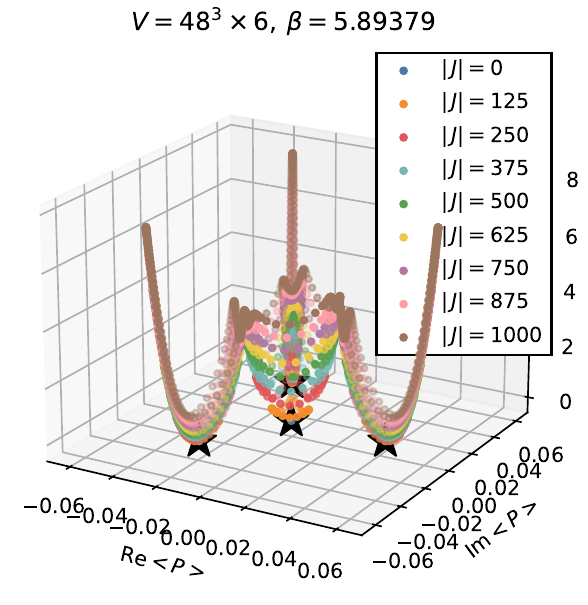}
\centering
   \includegraphics[width=5.3cm]
    {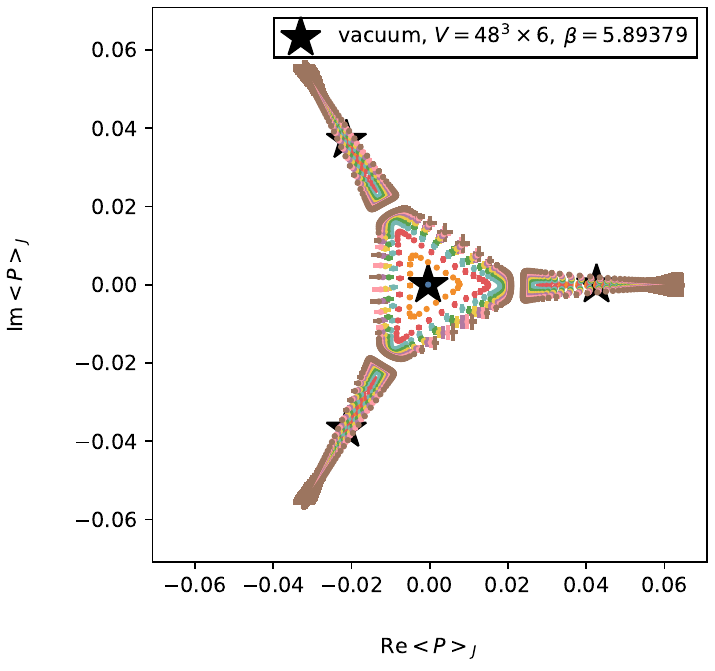}
    \caption{The effective potential (thermodynamic potential) $V_{\rm eff}(\langle P \rangle)$ of the Polyakov loop $\langle P \rangle
    ={\rm Re} \langle P \rangle+i~{\rm Im} \langle P \rangle
\in {\bf C}$ at the critical temperature $T_c$. 
    The horizontal value is multiplied 
    by the four-dimensional volume $N_s^3\times N_t$. 
    The color corresponds to the absolute value of the source $J$. 
Each dot corresponds to the adopted source $J=J_{\rm Re}+iJ_{\rm Im} \in {\bf C}$, and the total number of the dots is 901. 
The figure below is its 2D projection.
    }
    \label{fig:effective_pot}
\end{figure}

Figure~\ref{fig:effective_pot} shows 
a bird’s eye view of the effective potential  $V_{\rm eff}(\langle P \rangle)$ plotted against the Polyakov loop $\langle P \rangle
={\rm Re} \langle P \rangle+i~{\rm Im} \langle P \rangle
\in {\bf C}$
at the critical temperature $T_c$.
In this figure, the total number of the dots is 901, and 
each dot corresponds to the adopted source 
$J=J_{\rm Re}+iJ_{\rm Im} \in {\bf C}$.   
From Fig.~\ref{fig:effective_pot}, 
it can be seen that 
the effective potential $V_{\rm eff}[\langle P \rangle]$ 
is well depicted around the $Z_3$-symmetric 
and three $Z_3$-broken vacua
and also around the three lines between them. 

\section{Summary and Conclusion}

In the field-theoretical way, 
we have studied for the first time 
the effective potential of the Polyakov loop at finite temperature, i.e., the thermodynamic potential, in lattice QCD. 
In the framework of 
the reweighting method in lattice QCD, 
we have expressed the effective potential 
$V_{\rm eff}(\langle P \rangle)$
using the expectation value 
at the zero source of $J=0$.  

We have focused on the most difficult and interesting case 
of vacuum coexistence at the critical temperature $T_c$. 
We have adopted SU(3) quenched QCD on a spatially large lattice of $48^3 \times 6$ at 
$\beta$ = 5.89379, corresponding to the critical temperature $T_c$ of the deconfinement phase transition. 

Preparing a large number of 200,000 gauge configurations, 
we have numerically calculated the effective potential 
using the vacuum-associated reweighting method 
where gauge configurations around each vacuum are used separately, after categorizing them into each of the 
$Z_3$-symmetric and three $Z_3$-broken vacua.
The obtained Polyakov-loop effective potential 
$V_{\rm eff}(\langle P \rangle)$ is well depicted 
around the $Z_3$-symmetric and $Z_3$-broken vacua 
and also around the three lines between them. 
%
The lattice-QCD driven effective potential would be useful for modeling the Polyakov loop in hot QCD \cite{FS2017,FLLP2011}. 

The method present in this paper 
can be generally applied in lattice QCD. 
As a next step towards 
complete description of the QCD (phase) transition, 
it is useful to study the effective potential 
of the Polyakov loop 
at various temperatures in both quenched and full QCD.
In particular, the dynamical quark effect is quite 
interesting because it explicitly breaks 
the $Z_3$ symmetry \cite{BBDPSVW2016}.  
In fact, above $T_c$ in full QCD, 
one ground-state vacuum with a real Polyakov loop 
$\langle P \rangle \in {\bf R}$ is favored 
among the three $Z_3$-broken vacua, 
and the shape of the effective potential 
would be tilted, which quantitatively changes 
the appearance probability of each $Z_3$-broken vacuum. 
%
The Polyakov-loop fluctuation \cite{LFKRS2013} 
is also interesting to figure out 
the deconfinement transition, 
because it is related to 
the flatness of the Polyakov-loop effective potential 
around the minimum point.

\section*{Acknowledgments}
H.S. is supported by a Grants-in-Aid for 
Scientific Research [19K03869] from Japan Society 
for the Promotion of Science. 

\end{document}